\documentclass[letter,twocolumn]{jpsj3}
\usepackage{txfonts}

\usepackage{epsfig} 
\usepackage{subfigure}
\usepackage{bm}
\usepackage{color}
\usepackage{amsmath,amssymb}

\title{Magnetic Anisotropy by Rashba Spin-Orbit Coupling in Antiferromagnetic Thin Films}

\author{Jun'ichi Ieda$^1$\thanks{ieda.junichi@jaea.go.jp}, Stewart E. Barnes$^2$, and Sadamichi Maekawa$^1$}
\inst{
$^1$Advanced Science Research Center, Japan Atomic Energy Agency, Tokai 319-1195, Japan \\
$^2$Physics Department, University of Miami, Coral Gables, FL 33124, USA}

\abst{Magnetic anisotropy in an antiferromagnet (AFM) with inversion symmetry breaking (ISB) is investigated.
The magnetic anisotropy energy (MAE) resulting from the Rashba spin-orbit and  \emph{s}-\emph{d} type exchange interactions is determined for two different models of AFMs. The global ISB model, representing the effect of a surface, an interface, or a gating electric field, results in an easy-plane magnetic anisotropy.  
In contrast, for a local ISB model, i.e., for a noncentrosymmetric AFM, perpendicular magnetic anisotropy (PMA) arises. Both results differ from the ferromagnetic case, in which the result for PMA depends on the band structure and dimensionality. These MAE contributions play a key role in determining the direction of the N\'{e}el order parameter in antiferromagnetic nanostructures, and reflect the possibility of electrical-field control of the N\'{e}el vector. }

\begin{document}
\maketitle

Spin-orbitronics\cite{so_review} is a new trend in spin current physics\cite{book} that exploits the relativistic spin-orbit interaction in materials and opens fascinating new perspectives for both basic research and device technology.
A combination of spin-orbit interaction and the  \emph{s}-\emph{d} exchange interaction between the conduction electron spins and localized moment gives rise to a variety of phenomena such as the formation of skyrmions, spin-orbit torques, 
spin-charge conversion, magnetoresistance, and magnetic anisotropy. 
These advanced concepts and functionalities, originally recognized in ferromagnet (FM)-based nanohybrid structures, are also useful and even more salient in antiferromagnets (AFMs) because they offer pathways to manipulate AFMs, thereby fueling the recent development of antiferromagnetic spintronics\cite{afm_review,afm_review2}.

The magnetic anisotropy determines the energy barrier between the preferable orientations of (staggered) magnetization in (A)FMs. Understanding the magnetic anisotropy energy (MAE) in AFMs is therefore of fundamental importance when devising magnetic memory bits that are  reliably robust against any external (thermal, magnetic field, and electric current) noise\cite{pma_review}. 
It has also been pointed out that a large value of the MAE in AFMs is reflected in the exchange bias field\cite{UmeSakFuk06}, which is routinely used to fix the magnetization direction at the AFM/FM interface in current magnetic memory technology.

Several mechanisms are known to induce the MAE in AFMs.
The dipolar interaction among magnetic ions has been shown to explain the MAE in a series of corundum-type transition-metal oxides such as Cr${}_2$O${}_3$ \cite{TacNag58,ArtMurFon65}.
Strong perpendicular magnetic anisotropy (PMA) was reported recently at the Co$(111)/\alpha$-Cr${}_2$O${}_3(0001)$ interface and results in perpendicular exchange-biased interlayer coupling\cite{ShiOik12,NozShi17}.
The crystalline MAEs of manganese transition-metal alloys have been studied theoretically by first-principle calculations including the spin-orbit interaction\cite{UmeSakFuk06,ShiKhmMry10}. 
The anisotropic spin Hall effects\cite{SklZhaJun16} and spin-orbit torques\cite{WadHowZel16,FukZhaDut16} of such bimetallic AFMs have been  extensively studied.
Shape-induced MAE arises in compensated AFMs with strong magnetoelastic coupling, where it is analogous to the demagnetization energy in FMs\cite{GomLok07}.
A direction-dependent anisotropic exchange interaction seeds MAEs that can switch the preferred magnetization direction at the paramagnetic--ordered phase transition\cite{IshBal15}.

Here we focus on the effect of Rashba spin-orbit (RSO) interaction on the MAE in antiferromagnetic thin films. 
RSO coupling, which appears in a system with inversion symmetry breaking (ISB), plays a leading role in spintronics and other important branches of condensed matter physics\cite{rashba_review,rashba_review2}. 
For an RSO-coupled FM, we have derived the MAE\cite{BarIedMae14}, where the onset of PMA is explained by the energy gain from enhanced exchange splitting due to the RSO interaction. This is maximum when the magnetization is directed perpendicular to the ISB plane.
An important observation is that the induced MAE is quadratic in the RSO coupling constant, which  explains an even component of the electrical-field  modulation of the MAE in ferromagnetic thin films\cite{BarIedMae14,XuZha12,KimLeeLee16}.
In contrast, for an RSO-coupled AFM, we show below that the condition for PMA depends strongly on the type of RSO coupling, whereas the magnitude of the MAE shows the same quadratic dependence on the RSO coupling constant. 

\begin{figure}[b]
\begin{center}
\includegraphics[scale=0.7]{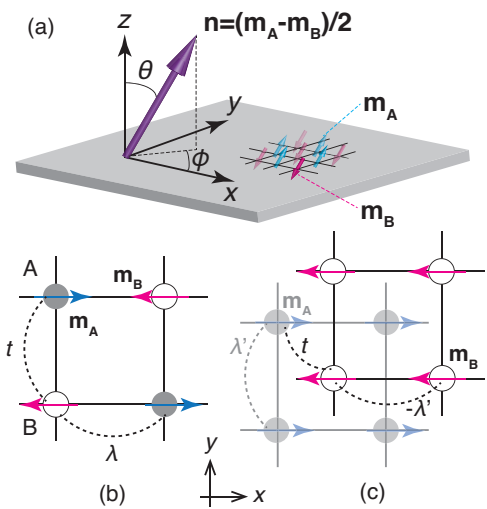}
\caption{(Color online) Schematic view of a system with the sublattice moments $\bm{m}_A$ and $\bm{m}_B$. 
(a) The direction of the N\'{e}el order parameter $\bm{n}$ is specified by the Euler angles $\theta$ and $\phi$. 
Inversion symmetry is broken along the $z$ axis, which is perpendicular to the film plane.
Models of a two-dimensional square lattice for nearest-electron hopping $t$ with (b) global Rashba coupling $\lambda$ and (c) local Rashba coupling $\pm\lambda'$ depending on the sublattices.} 
\label{fig1}
\end{center}
\end{figure}

%
%
To illustrate the effect of the RSO interaction on the MAE in antiferromagnetic thin films, we study two representative lattice models: a two-sublattice ordered AFM with global ISB or local ISB, as shown schematically in Fig.\ref{fig1}a. 
The former is a model of structural ISB at a surface or an interface\cite{ZelGaoVyb14,ZelGaoAur17}, whereas the latter is a model of a noncentrosymmetric AFM\cite{TanZhoXu16,SmeZelSin17}.

We start with the two-dimensional (2D) Rashba model introduced in Refs. [\citen{ZelGaoVyb14,ZelGaoAur17}] to simulate common experimental geometries in which a thin antiferromagnetic film is interfaced with another layer or subjected to a gating electric field.
We consider a square lattice AFM composed of two sublattices ($A$ and $B$) with equal saturation magnetization $M_\mathrm{S}$ and with a direction given by the classical unit vectors $\bm{m}_i$ for the $i$-th site. 
A uniform sublattice magnetization $\bm{m}_i=\bm{m}_{A/B}$ is assumed, as the $i$-th site belongs to the $A/B$ sublattice. 
(The spin dynamics due to nonuniform and time-dependent $\bm{m}_i$ in AFMs were studied previously in a continuous model\cite{afm-smf,afm-stt}.) 
The Hamiltonian 
\begin{align}\label{H}
H=\sum_{\langle ij\rangle}A_0 \bm{m}_i\cdot\bm{m}_j+H_0+\sum_{i}J_\mathrm{sd}\hat{\bm{s}}_i\cdot\bm{m}_i+H_\mathrm{R},
\end{align}
where the indices $i$, $j$ denote lattice sites; $\langle ij \rangle$ represents the sum of the nearest neighbors; $A_0>0$ is the antiferromagnetic exchange coupling constant between nearest-neighbor local moments; and $J_\mathrm{sd}$ is the on-site  \emph{s}-\emph{d} exchange coupling constant between the local moment and the conduction spin. Further, $\hat{\bm{s}}_i= c_i^\dagger \hat{\bm{\sigma}} c_i$, where $c_i^\dagger=(c_{i\uparrow}^\dagger,\, c_{i\downarrow}^\dagger)$ is the electron creation operator on the $i$-th site with spin $\uparrow$ or $\downarrow$, and $\hat{\bm{\sigma}}$ denotes the Pauli matrices. Here
$H_0$ represents the nearest-neighbor electron hopping, 
$H_0=-t\sum_{\langle ij\rangle}c^\dagger_ic_j$,
and $H_\mathrm{R}=H_\mathrm{R}^\mathrm{G}$ is the RSO coupling term, 
\begin{align}\label{RSO1}
H_\mathrm{R}^\mathrm{G}= i\lambda\sum_{\langle ij \rangle} \bm{\mu}_{ij}\cdot c_i^\dagger \hat{\bm{\sigma}} c_j,
\end{align} 
where $\lambda$ is the RSO constant (we set the lattice constant $a$ to 1), and $\bm{\mu}_{ij}$ $(=-\bm{\mu}_{ji})$ is the unit vector perpendicular to both the directions of hopping ($\bm{i}-\bm{j}$) and the ISB along the $z$ direction. The coupling constant $\lambda$ is proportional to the sum ($E_0 + E$) of the electric field $E_0$ reflecting the surface and that due to gating, $E$. 
 
The Hamiltonian (\ref{H}) is expressed as
$H=\sum_{\bm{k}}c^\dagger_{\bm{k}}\mathcal{H} c_{\bm{k}},$
in terms of the Fourier transforms $c_{\bm{k}}^\dagger=(c_{\bm{k}A\uparrow}^\dagger,\, c_{\bm{k}A\downarrow}^\dagger,c_{\bm{k}B\uparrow}^\dagger,\, c_{\bm{k}B\downarrow}^\dagger)$ of the $A$ and $B$ sublattice operators $c^\dagger_i$, and
\begin{align}\label{H1}
\mathcal{H}=\left[\gamma_{\bm{k}}-\lambda\left(\sin k_x \hat{\sigma}_y- \sin k_y\hat{\sigma}_x\right)\right]\hat{\tau}_x +J_\mathrm{sd}\bm{n}\cdot\hat{\bm{\sigma}}\hat{\tau}_z,
\end{align}
where $\gamma_{\bm{k}}=-2t(\cos k_x +\cos k_y)$, $\bm{n}=(\bm{m}_A-\bm{m}_B)/2$ is the N\'{e}el order parameter (Fig.\ref{fig1}a), and $\hat{\tau}_{x,z}$ are the Pauli matrices acting on the sublattice space. 
Here we assume a strong exchange $A_0$, so $\bm{m}_A=-\bm{m}_B$ (and $|\bm{n}|=1$). 
This assumption is valid even in the presence of inhomogeneous Dzyaloshinskii--Moriya interaction\cite{BogRosWol02}, which might be induced by a combination of the RSO interaction and on-site  \emph{s}-\emph{d} exchange interaction\cite{ImaBruUts04,KunZha15}.

Using the Pauli matrix algebra on $\mathcal{H}^2$ and $[\mathcal{H}^2- (\gamma_{\bm{k}}^2+J_\mathrm{sd}^2+\lambda^2\kappa^2_{\bm{k}})]^2$ gives four 
 energy eigenvalues of Eq.(\ref{H1}):
\begin{align}\label{eigen1}
\epsilon_{\bm{k}\eta s}(\bm{n})=\eta\sqrt{\gamma_{\bm{k}}^2+J_\mathrm{sd}^2+\lambda^2\kappa^2_{\bm{k}}
-2s \lambda \kappa_{\bm{k}} S_{\bm{k}}(\bm{n})},
\end{align}
where we define $\kappa_{\bm{k}}=(\sin^2k_x+\sin^2k_y)^{1/2}$, and
$S_{\bm{k}}(\bm{n})=\sqrt{\gamma_{\bm{k}}^2+J_\mathrm{sd}^2
\left[1-\sin^2\theta\sin^2(\phi_{\bm{k}}-\phi)
\right]},$
with $\bm{n}=(\sin\theta\cos\phi,\sin\theta\sin\phi,\cos\theta)$, as shown in Fig.\ref{fig1}a, and $\tan\phi_{\bm{k}}=\sin k_y/\sin k_x$. 
The eigenvalues with indices $\eta=\pm1$ and $s=\pm1$ are identified as conduction/valence and minority/majority-spin bands, respectively. 
The square root
$S_{\bm{k}}(\bm{n})$ is a decreasing function of $\sin\theta$, and the magnitude of the spin splitting is maximum for $\theta=0$, at which
the eigenvalues become independent of $\phi_{\bm{k}}$ as $\epsilon_{\bm{k}\eta s}(\hat{\bm{z}})=\eta\left|\sqrt{\gamma_{\bm{k}}^2+J_\mathrm{sd}^2}-s\lambda\kappa_{\bm{k}}\right|.$
For the $\bm{k}$ points with $\kappa_{\bm{k}}=0$, band crossing occurs owing to $\mathcal{PT}$ symmetry, where $\mathcal{P}$ and $\mathcal{T}$ represent the inversion and time-reversal symmetries, respectively, which are broken individually in Eq.(\ref{H1}). 

The MAE is defined as the difference in the sums over occupied  states of eigenvalues (\ref{eigen1}) with $\bm{n}=\hat{\bm{z}}$ as the reference,
\begin{align}\label{MAE1}
E_\mathrm{MAE}= \sum_{\bm{k}\eta s}^\mathrm{occ.}\epsilon_{\bm{k}\eta s}(\bm{n})
-\sum_{\bm{k}\eta s}^\mathrm{occ.}\epsilon_{\bm{k}\eta s}(\hat{\bm{z}}).
\end{align}
Expanding Eq.(\ref{MAE1}) around $\theta\sim0$ 
yields
\begin{align}
E_\mathrm{MAE}= K\sin^2\theta,
\end{align}
where the uniaxial magnetic anisotropy constant $K$ is given by
\begin{align}\label{K1}
K 
= \sum_{\bm{k}\eta s}^\mathrm{occ.}
 \frac{\eta s J_\mathrm{sd}^2\lambda\kappa_{\bm{k}}\sin^2(\phi_{\bm{k}}-\phi)}{2\sqrt{\gamma_{\bm{k}}^2+J_\mathrm{sd}^2}\left|\sqrt{\gamma_{\bm{k}}^2+J_\mathrm{sd}^2}-s\lambda\kappa_{\bm{k}}\right|}. 
\end{align}
This is one of the main results in this paper.

The sign of $K$ determines the type of MAE: PMA ($K>0$) or easy-plane anisotropy ($K<0$).
First, we consider $\gamma_{\bm{k}}^2+J_\mathrm{sd}^2-\lambda^2\kappa_{\bm{k}}^2>0$ for all the $\bm{k}$ points; i.e., band inversion due to the RSO interaction does not occur. 
Without loss of generality, we assume $\lambda>0$.
Then the above condition can be expressed as $\lambda < \lambda_\mathrm{c}$,
where the critical value, evaluated at the band touching points, $\bm{k}=(\pi/2,\pm\pi/2)$, $(-\pi/2,\pm\pi/2)$, is 
$\lambda_\mathrm{c}=J_\mathrm{sd}/\sqrt{2}$ for the current model.
After the spin summation in Eq.(\ref{K1}) we have $K= \sum_{\bm{k}\eta}^\mathrm{occ.} \eta f(\bm{k})$ for $\lambda < \lambda_\mathrm{c}$, with
\begin{align}\label{f}
f(\bm{k}) 
 = \frac{  J_\mathrm{sd}^2\lambda^2\kappa_{\bm{k}}^2\sin^2(\phi_{\bm{k}}-\phi)}{\sqrt{\gamma_{\bm{k}}^2+J_\mathrm{sd}^2}
 \left|\gamma_{\bm{k}}^2+J_\mathrm{sd}^2-\lambda^2\kappa_{\bm{k}}^2\right|} >0.
\end{align}
From this, we observe that the valence band ($\eta=-1$) makes a negative contribution to $K$, whereas 
that of the conduction band is reversed as $\eta$ changes sign. 
In total, for partially occupied energy bands, it follows that $K<0$; i.e., the RSO-induced MAE for the N\'{e}el order parameter $\bm{n}$ would be the easy-plane type within the model (\ref{H}).
It becomes maximum when the band is half-filled (the only $\eta=-1$ band is fully occupied).
We remark that the MAE arises from a combination of the RSO and  \emph{s}-\emph{d} exchange interactions, both of which are crucial factors for spin splitting of the energy bands (\ref{eigen1}). 

\begin{figure}[b]
\begin{center}
\includegraphics[scale=0.8]{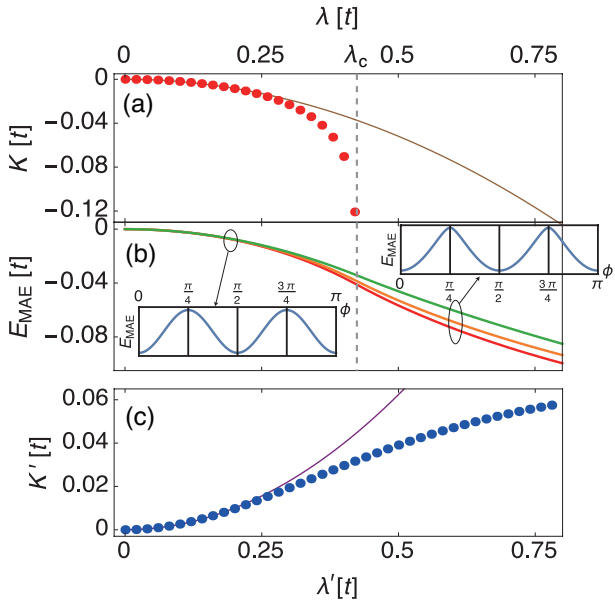}
\caption{(Color online) Rashba coupling dependence of the MAE for the half-filled band. 
(a) Dots represent the magnetic anisotropy constant $K$ in Eq.(\ref{K1}) for the global RSO model. The solid curve and vertical dashed line indicate the parabolic fit to the data and the critical value, $\lambda=\lambda_\mathrm{c}$, defined in the text, respectively. 
(b) MAE $E_\mathrm{MAE}$ for the global RSO model, Eq.(\ref{MAE1}), for $\theta=\pi/2$ with $\phi=0$, $\pi/8$, and $\pi/4$ (from bottom to top).
The insets show the $\phi$ dependence of $E_\mathrm{MAE}$ for $\lambda=0.2$ (left) and 0.6 (right), as indicated by the arrows.
(c) The dots represent the magnetic anisotropy constant $K'$ in Eq.(\ref{K2}) for the sublattice-dependent RSO model. The solid curve indicates the parabolic fit to the data.
The energy unit is $t$, and we use $J_\mathrm{sd}=0.6$ for all the plots.} 
\label{fig2}
\end{center}
\end{figure}

Figure \ref{fig2}a shows the RSO coupling dependence of $K$ for the half-filled band.
For small values of $\lambda$, the anisotropy constant $K$ in Eq.(\ref{K1}) is proportional to the squared RSO coupling, $\lambda^2$, as in the ferromagnetic case\cite{BarIedMae14}.  
When $\lambda \to \lambda_\mathrm{c}$, on the other hand, $K$ grows rapidly, as expected from the denominator of $f(\bm{k})$.
When $\lambda > \lambda_\mathrm{c}$, at which band inversion occurs, i.e., $\gamma_{\bm{k}}^2+J_\mathrm{sd} ^2<\lambda^2\kappa_{\bm{k}}^2$ holds for certain pockets of $\bm{k}$,
Eq.(\ref{K1}) diverges, implying that the expansion of $E_\mathrm{MAE}$ with $\sin^2\theta$ becomes invalid.
In fact, for $\lambda = \lambda_\mathrm{c}$, a $|\sin\theta|$ component appears in the expansion of $E_\mathrm{MAE}$.
To avoid this complexity, we compute Eq.(\ref{MAE1}) directly, as shown in Fig.\ref{fig2}b, for $\theta=\pi/2$ with $\phi=0$ ($\bm{n}||[100]$), $\pi/8$, and $\pi/4$ ($\bm{n}||[110]$).
$E_\mathrm{MAE}$ is continuous and has an inflection point at $\lambda = \lambda_\mathrm{c}$.
In the inset of Fig.\ref{fig2}b, we show the $\phi$ dependence of $E_\mathrm{MAE}$, which reveals that the $[100]$ and $[010]$ directions are equivalent easy directions. 
We also observe that the in-plane MAE is well-fitted by $\sin^2(2\phi)$ for $\lambda < \lambda_\mathrm{c}$, whereas it becomes anharmonic for $\lambda > \lambda_\mathrm{c}$.

Next, we consider a simple 2D model of noncentrosymmetric AFMs that has been introduced to study current-induced manipulation of Dirac fermions by spin-orbit torques in CuMnAs\cite{SmeZelSin17}. 
 The tetragonal CuMnAs lattice, which can be stabilized by molecular beam epitaxial growth on GaAs or GaP\cite{WadNovCam13}, is inversion-symmetric, whereas its Mn spin sublattices form noncentrosymmetric inversion partners, supporting the presence of the staggered RSO interaction for intrasublattice electron hopping.
To model this system, we consider a tetragonal primitive structure with a bipartite square lattice in the $x$--$y$ plane whose sublattice atoms, labeled \emph{A} and \emph{B}, are buckled in the $z$ direction, as shown in Fig.\ref{fig1}c.
(We omit here the next-nearest hopping $t'$ and the dispersion along the $z$ direction for simplicity.)
The RSO interaction adopted here is defined for hopping between the same sublattice sites and changes sign depending on the sublattice:
\begin{align}\label{RSO2}
H_\mathrm{R}^\mathrm{L}= i\lambda'\sum_{\langle\langle ij \rangle\rangle} (-1)^i\bm{\mu}_{ij}\cdot c_i^\dagger \hat{\bm{\sigma}} c_j,
\end{align} 
where $\langle\langle ij \rangle\rangle$ denotes the next-nearest neighbors.
The Hamiltonian (\ref{H}) is now
$H'=\sum_{\bm{k}}c^\dagger_{\bm{k}}\mathcal{H}' c_{\bm{k}},$
with 
\begin{align}\label{H2}
\mathcal{H}'=\gamma'_{\bm{k}}\hat{\tau}_x -\left[\lambda'\left(\sin k_x \hat{\sigma}_y- \sin k_y\hat{\sigma}_x\right)-J_\mathrm{sd}\bm{n}\cdot\hat{\bm{\sigma}}\right]\hat{\tau}_z,
\end{align}
where $\gamma'_{\bm{k}}=-2t\cos (k_x/2)\cos (k_y/2)$.
The eigenvalues
\begin{align}\label{eigen2}
\epsilon'_{\bm{k}\eta}(\bm{n})=\eta\sqrt{{\gamma'_{\bm{k}}}^2+J_\mathrm{sd}^2+\lambda'^2\kappa^2_{\bm{k}}
-2 \lambda' \kappa_{\bm{k}} S'_{\bm{k}}(\bm{n})},
\end{align}
with
$S'_{\bm{k}}(\bm{n})=J_\mathrm{sd}\sin\theta\sin\left(\phi_{\bm{k}}-\phi\right)$,
 are doubly degenerate for all the $\bm{k}$ points owing to $\mathcal{PT}$ symmetry.

The MAE is defined similarly to Eq.(\ref{MAE1}):
\begin{align}\label{MAE2}
E'_\mathrm{MAE}= 2\sum_{\bm{k}\eta}^\mathrm{occ.}\epsilon'_{\bm{k}\eta}(\bm{n})
-2\sum_{\bm{k}\eta}^\mathrm{occ.}\epsilon'_{\bm{k}\eta}(\hat{\bm{z}}),
\end{align}
where the factor of 2 comes from the $\mathcal{PT}$ degeneracy. 
Using Eq.(\ref{eigen2}), 
we expand Eq.(\ref{MAE2}) around $\theta\sim0$ and obtain
\begin{align}
E'_\mathrm{MAE}= K'\sin^2\theta,
\end{align}
where
\begin{align}\label{K2}
K' = -\sum_{\bm{k}\eta}^\mathrm{occ.}
 \frac{\eta J_\mathrm{sd}^2\lambda'^2\kappa_{\bm{k}}^2\sin^2(\phi_{\bm{k}}-\phi)}{({\gamma'_{\bm{k}}}^2+J_\mathrm{sd}^2+\lambda'^2\kappa_{\bm{k}}^2)^{3/2}}.
\end{align}
This is the second key result of this paper.
The linear term in $\lambda'$ appears in the expansion of Eq.(\ref{eigen2}), but it vanishes after the $\bm{k}$ summation owing to the oddness of the directional factor $\sin\left(\phi_{\bm{k}}-\phi\right)$.
It is obvious that $K'>0$ for the partially occupied energy bands; i.e., PMA is always favored in the system with sublattice-dependent RSO coupling (\ref{RSO2}). 
Figure \ref{fig2}c shows the RSO coupling dependence of $K'$ for the half-filled band.
For small values of $\lambda'$, the anisotropy constant $K'$ is proportional to the squared RSO coupling, $\lambda'^2$, and it deviates from a parabola owing to the denominator in Eq.(\ref{K2}). 
We note that Kim \emph{et al.} numerically studied a similar system\cite{KimKanSch} that supports the general tendency of our analytical result [Eq.(\ref{K2})]. 

It has been pointed out\cite{TanZhoXu16,SmeZelSin17} that when the N\'{e}el vector is along the $[100]$ or $[010]$ direction, the energy bands (\ref{eigen2}) possess two Dirac points, 
where the fourfold band degeneracy is protected by the glide mirror plane symmetry in addition to the $\mathcal{PT}$ symmetry. 
Once the N\'{e}el vector has a $z$ component, these Dirac points are gapped, resulting in reduction of the total band energy. 
The  \emph{s}-\emph{d} exchange  field along the $z$ direction plays a role similar to that of the perpendicular magnetic field on the ordinary twofold Dirac point.
This is the physical picture of the PMA scenario for the Dirac AFM system reflected in the present model.

%
%


Nanostructured AFMs exhibit a shape-induced MAE\cite{GomLok07} that causes the orientation of the N\'{e}el vector to align with the surface/interface plane.   
For example, antiferromagnetic spin structure in tetragonal CuMnAs was investigated by a combination of neutron diffraction and X-ray magnetic linear dichroism (XMLD) measurements\cite{WadHilSha15}. These measurements imply an easy-plane MAE. 
The authors of Ref. [\citen{WadHilSha15}] argue that their neutron data, supplemented by \emph{ab initio} calculations, imply that the Mn spins are confined in the $(ab)$ plane.
Recent XMLD microscopy imaging of a tetragonal CuMnAs film reveals an inhomogeneous domain structure at the submicron level\cite{GrzWadEdm17}. 
The observed complex multidomain structure implies the influence of a destabilizing factor on the in-plane spin textures.
The RSO-induced PMA described in this work can be considered as part of that scenario.


In conclusion, we showed that the RSO interaction produces the MAE for two-sublattice AFMs with broken inversion symmetry.
Two types of the Rashba coupling were considered.
With regard to the Rashba coupling defined for hopping between different sublattice sites, the uniaxial magnetic anisotropy constant becomes negative for weak Rashba coupling, and biaxial in-plane easy axes are identified. 
This Rashba model is appropriate for a common geometry for an antiferromagnetic thin film and other materials with hybrid structures, where it is possible to modulate the RSO coupling by attaching a nonmagnetic film to an AFM, or, more directly, by electric field gating.
In contrast, for the Rashba coupling defined for hopping between the same sublattice sites, PMA is favored, and a band gap due to the \emph{s}-\emph{d} exchange interaction appears.
This feature is a potential obstacle to realization of a Dirac AFM, as recently proposed\cite{TanZhoXu16,SmeZelSin17} for a similar model system, because it requires the in-plane N\'{e}el vector configuration.
Although further investigation is needed to apply our simple model study to realistic systems,
our finding offers a way to tune the magnitude of the MAE by a suitable choice of material combinations and by electrical gating.

\begin{acknowledgments}
The authors thank S. Shamoto for a fruitful discussion on neutron diffraction experiments and J. Sinova and O. Gomonay for valuable comments.
This research was supported by JSPS KAKENHI (JP16K05424, JP17H02927, JP26103006) and by the Exploratory Research for Advanced Technology (ERATO) program of the Japan Science and Technology Agency (JST) of MEXT, Japan.
\end{acknowledgments}

\end{document}